\documentclass[fleqn,usenatbib]{mnras}

\usepackage {psfig}
\usepackage{amssymb}
\usepackage {graphicx}
\usepackage {epsfig}
\usepackage{mathrsfs}
\usepackage[english]{babel}
\usepackage{amsfonts}
\usepackage{fontenc}
\usepackage{textcomp}
\usepackage{microtype}
\usepackage[pdftex]{lscape}
\usepackage{amsmath}

\title[Chaos in $N$-body simulations]{$N$-body chaos and the continuum limit in numerical simulations of self-gravitating systems, revisited}
\author[Di Cintio and Casetti]{Pierfrancesco Di Cintio$^{1,2,3}$\thanks{E-mail:p.dicintio@ifac.cnr.it} and Lapo Casetti$^{3,2,4}$\\
$^{1}$Consiglio Nazionale delle Ricerche, Istituto di Fisica Applicata ``Nello Carrara",\\
 via Madonna del piano 10, I-50019 Sesto Fiorentino, Italy\\
$^{2}$INFN -  Sezione di Firenze, via G.\ Sansone 1, I-50019 Sesto Fiorentino, Italy\\
$^3$Dipartimento di Fisica e Astronomia, Universit\`a di Firenze, via G.\ Sansone 1, I-50019 Sesto Fiorentino, Italy\\
$^4$ INAF - Osservatorio Astrofisico di Arcetri, largo Enrico Fermi 5, I-50125 Firenze, Italy}
\hypersetup{draft}
\begin{document}
\date{Accepted 2019 September 5. Received 2019 September 4; in original form 2019 January 28}
\pubyear{0000}
\maketitle
\begin{abstract}
We revisit the r\^{o}le of discreteness and chaos in the dynamics of self-gravitating systems by means of $N$-body simulations with active and frozen potentials, starting from spherically symmetric stationary states and considering the orbits of single particles  in a frozen $N$-body potential as well as the orbits of the system in the full $6N$-dimensional phase space. We also consider the intermediate case where a test particle moves in the field generated by $N$ non-interacting particles, which in turn move in a static smooth potential. We investigate the dependence on $N$ and on the softening length of the largest Lyapunov exponent both of single particle orbits and of the full $N$-body system. For single orbits we also study the dependence on the angular momentum and on the energy. Our results confirm the expectation that orbital properties of single orbits in finite-$N$ systems approach those of orbits in smooth potentials in the continuum limit $N \to \infty$ and that the largest Lyapunov exponent of the full $N$-body system does decrease with $N$, for sufficiently large systems with finite softening length. However, single orbits in frozen models and active self-consistent models have different largest Lyapunov exponents and the $N$-dependence of the values in non-trivial, so that the use of frozen $N$-body potentials to gain information on large-$N$ systems or on the continuum limit may be misleading in certain cases.  
\end{abstract}
\begin{keywords}
Chaos -- gravitation -- galaxies: evolution -- methods: numerical 
\end{keywords}

\section{Introduction}
The dynamics of $N$-body self-gravitating systems, due to the long-range nature of the $1/r^2$ force, is dominated by mean field effects rather than by inter-particle collisions, for sufficiently large particle number $N$. It is then natural to adopt a description in the continuum and collisionless limit $N\to\infty$, with $m\to 0$, where $m$ is the individual particles' mass. The system is described by the single-particle distribution function $f(\mathbf{r},\mathbf{v},t)$ in phase space, where $\mathbf{r}$ is the position and $\mathbf{v}$ is the velocity, and the time evolution of $f$ is dictated by the collisionless Boltzmann-Poisson equations  (CBE, see e.g.\ \citealt{2008gady.book.....B}) 
\begin{equation}\label{vlasovpoisson}
\begin{cases}
\displaystyle \partial_t f+\mathbf{v}\cdot\nabla_{\mathbf{r}}f+\nabla\Phi\cdot\nabla_{\mathbf{v}}f=0\\
\displaystyle \Delta\Phi(\mathbf{r})=-4\pi G\rho(\mathbf{r})=-4\pi G\int f\, {\rm d}\mathbf{v},
\end{cases}
\end{equation} 
linking $f$ to the (in principle time-dependent) density-potential pair $(\rho,\Phi)$; $G$ is the gravitational constant. Equations \eqref{vlasovpoisson} yield a faithful description of the dynamics as long as the effect of binary encounters on the time evolution of $f$ may be neglected: this happens for times shorter than the two-body relaxation time $t_{2b}$. The relevant fact (holding true for any long-range-interacting system, not only for self-gravitating ones; see e.g.\ \citealt{CampaEtAl:book,CampaEtAl:physrep}) is that such a timescale grows with $N$. In the case of self-gravitating systems the relaxation time may be estimated as (\citealt{1941ApJ....93..285C,2008gady.book.....B})
\begin{equation}\label{t2b}
t_{2b}=\frac{v^3_{\rm typ}}{8\pi(Gm)^2n\ln\Lambda}\approx \frac{N}{8\ln N} t_{\rm dyn},
\end{equation}
where $t_{\rm dyn}={r_c}/{v_{\rm typ}}$,
with $r_c$ and $v_{\rm typ}$ the typical size and velocity scale of the system, $n$ is an average number density and $\ln\Lambda$ is the so-called Coulomb logarithm, i.e., the logarithm of the ratio of the maximum to minimum impact parameter between the system's particles, typically a number of order 10. Therefore, large $N$ systems such as elliptical galaxies, where $N\approx 10^{11}$, have two-body relaxation times far exceeding the Hubble time, so that discreteness effects and dynamical collisions can be safely neglected, assuming Eqs.\ (\ref{vlasovpoisson}) valid over all physically relevant times.\\
\indent However, self-gravitating systems are often studied by means of $N$-body simulations, numerically integrating the equations of motion of a system of $N$ particles interacting via gravitational forces. First of all, we note that the in a numerical simulation the number of particles is much smaller than the real number of stars in a galaxy. As a consequence, the relaxation time $t_{2b}$, measured in units of the dynamical time $t_{\rm dyn}$, is orders of magnitudes smaller than the relaxation time of the system one would like to simulate\footnote{For instance, consider a galaxy of $10^{11}$ solar masses for which $t_{\rm dyn}\approx 10^{8}$ yrs and $t_{2b}\approx 10^{16}$ yrs. The relaxation time $t_{2b}$ of a \textit{direct} numerical simulation of the full $N$-body problem using $10^5$ particles would be much smaller, roughly by a factor $10^{-6}$!  Somewhat larger values of $N$, but still far from those of a typical galaxy (and we are referring only to stars: if we wold like to consider also the dynamics of dark matter particles the numbers would be much larger), may be reached with tree codes; particle-in-cell or particle-mesh simulations, allowing to simulate even larger systems, do not account for dynamical collisions and therefore for the latter the concept of ``collisional relaxation'' becomes somewhat unclear.}. Therefore, a direct numerical simulation of a collisionless system might be considered\footnote{The softening of the interactions at small distance might help in making the systems ``less collisional'': see Sec.\ \ref{sec_nummethods} below.} ``truly collisionless'' only for sufficiently small times.  Moreover, a self-gravitating (finite) $N$-body system is always a chaotic dynamical system, that is, its trajectories are always linearly unstable in phase space; however, its macroscopic properties are expected to be less and less sensitive to such a local instability for increasing $N$, if a continuum limit if reached when $N \to \infty$ .\\
\indent From the point of view of the numerical experiments, since the pioneering work by \cite{1964ApJ...140..250M,1971JCoPh...8..449M} (see also \citealt{2002ApJ...580..606H,2007arXiv0710.0514H}), an apparent contradiction has shown up: on the one hand, when increasing the number of particles of a model, while keeping for example the total mass fixed, the sensitivity to the initial conditions (and therefore the ``amount" of dynamical chaos) should increase, while on the other hand, its continuum limit represented by Eqs.\ \eqref{vlasovpoisson} is a non-canonical infinite-dimensional Hamiltonian system that admits an infinite number of conserved quantities (the so-called Casimir invariants, or casimirs, \citealt{1998ApJ...500..120K}) and one would expect the systems to become less chaotic when getting closer to the continuum limit, i.e., by increasing $N$. Technically speaking, the evolution under the CBE is invariant under the continuous group of particles re-labelling and thus is associated to infinite integral invariants, following Noether theorem. Such group is instead discrete for pure $N$-body systems, see e.g.\ \cite{2014EPJD...68..218E,2018RvMPP...2....9E}.\\
\indent All these observations led to question the validity of the continuum limit, at least from the point of view of numerical simulations and their interpretation (see e.g. \citealt{1998NYASA.848...28K,2018arXiv180406920E}).      
For example, in a series of papers, Kandrup and collaborators investigated the validity of such limit and the effect of discreteness noise using single particle orbit analysis in frozen $N$-body models (\citealt{2001PhRvE..64e6209K,2002PhRvE..65f6203S,2002PhDT........25S,2003ApJ...585..244K,2004PhRvS...7a4202K,2004CeMDA..90..147S}), and Langevin-like sochastic equations (\citealt{2001MNRAS.323..681K,2003MNRAS.341..927K,2003astro.ph.12434T,2003MNRAS.345..727K,2004CeMDA..88....1K,2004ApJ...602..678S}).\\ 
\indent Dynamical chaos, i.e., the exponential sensitivity to the initial conditions, is customarily measured by the largest Lyapunov exponent $\lambda_{\rm max}$ (see e.g.\ \cite{1992rcd..book.....L} and the discussion in Sec.\ \ref{sec_nummethods} below). In the context of one-dimensional gravity, \cite{2000chun.proc..259T,2000PhRvE..61..948T} investigated the dependence of $\lambda_{\rm max}$ on $N$ finding a curious $\lambda_{\rm max}\propto N^{-1/5}$ scaling. \cite{2011TTSP...40..360M} and \cite{2011PhRvE..84f6211G} repeated the analysis of the behaviour of $\lambda_{\rm max}$ with $N$ for another long-range-interacting one-dimensional system, the so-called Hamiltonian Mean-Field model (HMF, \citealt{1995PhRvE..52.2361A}) and its two-dimensional generalization. These studies suggest a scaling of the form $\lambda_{\rm max}\propto N^{-1/3}$ or $\lambda_{\rm max}\propto 1/\ln N$, depending on the total energy\footnote{Due to the long-range nature of the interaction and the fact that in numerical simulations the total energy is rescaled to a constant, all  trends with $N$ are valid also with the specific energy $E/N$ if total energy $E$ is not rescaled.} of the model in the finite $N$ regime; these results have been confirmed by \cite{2018JSMTE..03.3204F}. Similar scalings of $\lambda_{\rm max}$ with $N$ were found by \cite{2012PhRvE..86d1136M} for the self-gravitating ring model, where particles interacting via softened gravitational forces are constrained on a ring. Using a differential geometry approach (see e.g.\ \citealt{1996PhRvE..54.5969C,physrep2000} for a detailed review), \cite{1986A&A...160..203G} (see also \citealt{2009A&A...505..625G}) predicted that the exponential instability time scale (associated to the reciprocal of the Largest Lyapunov exponent) for a three dimensional self-gravitating $N$-body system grows as $t_{\rm exp}\propto N^{1/3}$ in units of a typical crossing time $t_{\rm cr}$. This latter result however, received strong criticism (see e.g.\ \citealt{1993NYASA.706...81K} and references therein, see also the discussion in \citealt{1995A&AT....7..225K}), as it appears to be based on a underestimation of the characteristic curvature of the Riemannian manifold.\\
\indent The numerical study of chaotic dynamics and Lyapunov exponents in full $N$-body self-gravitating systems in three dimensions has been carried out by \cite{1995PhRvE..51...53C} and \cite{2003Ap&SS.283..347C}, although for rather small systems, detecting a weak decrease of $\lambda_{\rm max}$ with increasing $N$, and by \cite{2002MNRAS.331...23E} who considered larger $N$'s and also detected a decreasing Lyapunov exponent when increasing $N$. In parallel, \cite{2002ApJ...580..606H} explored the linear stability of self-consistent equilibrium $N$-body systems as function of $N$ (for $N$ up to $\approx 10^5$), by computing the expansion of the spatial part of the cartesian variational vector, finding instead that the associated growth rate (slightly) increases with $N$, suggesting an analogous trend for the parent finite time Lyapunov exponent.\\
\indent More recently, \cite{2015MNRAS.449.4458S}, \cite{2018MNRAS.473.2348B} and \cite{2018arXiv180707084G} studied the effect of finite $N$ fluctuations on the properties of the end states of dissipationless cosmological simulations (see also \citealt{2004MNRAS.348...62R,2007PhRvD..75f3516J,2007PhRvD..76j3505J,2009MNRAS.398.1279S}) finding that, for analogous initial conditions, the particle number $N$ influences both the spurious collisional effects, and the onset of collective instabilities, associated to the long-range nature of the Newtonian force\footnote{Remarkably, even in the context of non-neutral plasmas finite-$N$ effects are observed in the asymptotic energy distribution of expanding spherical and ellipsoidal ion bunches, with respect to its continuum limit counterparts (\citealt{2011PhRvE..84e6404G,2013PhRvL.110m3401S,2014arXiv1408.3857D,2018PhRvS..21f4201Z})}.\\
\indent In this paper we revisit the problem of finite $N$-body chaos in equilibrium models of self-gravitating systems, considering the dynamics in the full $6N$-dimensional phase space as well as individual tracer particle orbits in self-consistent simulations and frozen $N$-body models. Our aim is to give a contribution to the clarification of some of the above mentioned issues, and in particular to understand whether the amount of chaos as measured by the largest Lyapunov exponent does decrease or not when an $N$-body self-gravitating system approaches its continuum limit, and whether a meaningful comparison is possible between single-orbit properties in frozen or self-consistent potentials and those of the full $N$-body dynamics, again when $N$ grows towards the continuum limit. The rest of the paper is structured as follows: in Section \ref{sec:model} we introduce the models and the tools to set the stage for the numerical simulations whose results are presented  and discussed in Section \ref{sec:simulations}; in Section \ref{summary} we summarize our findings. 
\section{Models and methods}\label{sec:model}
\subsection{Initial conditions}
We have performed two main types of numerical simulations: self-consistent $N$-body simulations of \textit{equilibrium} spherical systems and single particle orbit integrations in frozen $N$-body potentials, for various numbers of particles $N$. In addition, we have also performed auxiliary numerical experiments where a test particle is propagated in the (time-dependent) field of $N$ non-interacting particles moving a static smooth potential. In all cases, that is, frozen $N$-body, self-consistent models, and smooth orbits systems, the initial particle positions $\mathbf{r}_{i,0}$ are sampled from two different spherically symmetric density profiles, the flat cored \cite{1911MNRAS..71..460P} profile
\begin{equation}
\label{plum}
\rho(r)=\frac{3Mr_c^2}{4\pi\left(r^2+r_c^2\right)^{5/2}};
\end{equation}
and the cuspy \cite{1990ApJ...356..359H} profile
\begin{equation}\label{hernquist}
\rho(r)=\frac{M}{2\pi r_c^2r(1+r/r_c)^3},
\end{equation}
where  $r_c$ is a length scale.
In the continuum limit, the smooth potentials $\Phi(r)$ generated by the distributions (\ref{plum}) and (\ref{hernquist}) are central and integrable, thus admitting only regular orbits with $\lambda_{\rm max}=0$. For the equilibrium self-consistent models, in order to generate the velocities, we use the standard rejection technique to sample the modulus of the initial velocities $v_{i,0}$ from the isotropic equilibrium phase-space distribution function $f(\mathcal{E})$, with single particle energy per unit mass $\mathcal{E}=v^2/2+\Phi(r)$, obtained from $\rho$ by means of the \cite{1916MNRAS..76..572E} inversion of Eqs.\ (\ref{vlasovpoisson}) as
\begin{equation}\label{OM}
f(\mathcal{E})=\frac{1}{\sqrt{8}\pi^2}\frac{\rm d}{{\rm d}\mathcal{E}}\int_\mathcal{E}^{0}\frac{{\rm d}\rho}{{\rm d}\Phi}\frac{{\rm d}\Phi}{\sqrt{\Phi-\mathcal{E}}}.
\end{equation}
The direction of the velocity vectors is then assigned randomly by sampling a homogeneous distribution in the angular variables $(\vartheta,\varphi)$ and converting the result to Cartesian coordinates.
\begin{figure*}
\includegraphics[width=0.8\textwidth]{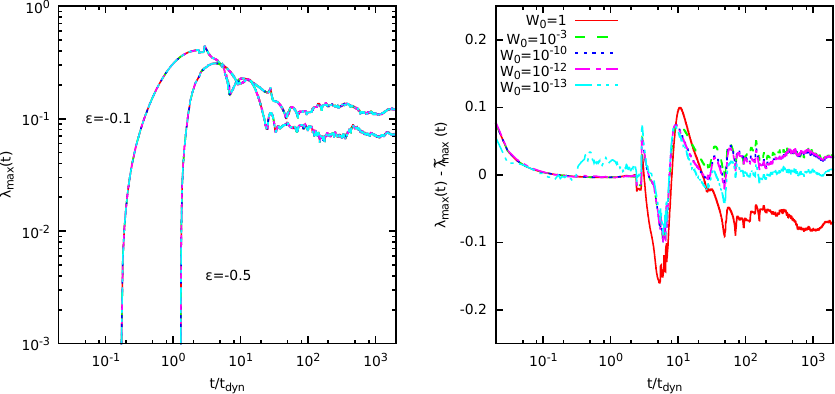}
\caption{Left panel: Evolution of $\lambda_{\rm max}(t)$ for two orbits with $\mathcal{E}=-0.1$ and $-0.5$ evolved in a frozen Plummer potential generated by $N = 10^5$ particles for different initial normalizations of the tangent dynamics $\mathbf{W}_{6}$. Right panel: Evolution of the difference $\lambda_{\rm max}(t)-\tilde{\lambda}_{\rm max}(t)$ for the case with $\mathcal{E}=-0.1$ and same choices of the initial normalization.}
\label{test1}
\end{figure*}
\subsection{Numerical methods and tests}
\label{sec_nummethods}
We use an adaptive order symplectic integrator (\citealt{1991CeMDA..50...59K,1995PhyS...51...29C}) with fixed time-step $\Delta t$ to solve the equations of motion 
\begin{equation}\label{eom}
\ddot{{\mathbf r}}_i=-Gm\sum_{j=1}^N\frac{{\mathbf r}_i-{\mathbf r}_j}{||{\mathbf r}_i-{\mathbf r}_j||^3},
\end{equation}
and their associated variational equations for the tangent vectors $\mathbf{w}_i$ (\citealt{1971JCoPh...8..449M,1993ApJ...415..715G,2002ApJ...580..606H,2016MNRAS.459.2275R})
\begin{equation}\label{var}
\ddot{{\mathbf w}}_i=-Gm\sum_{j=1}^N\left[\frac{{\mathbf w}_i-{\mathbf w}_j}{||{\mathbf r}_i-{\mathbf r}_j||^3}-3({\mathbf r}_i-{\mathbf r}_j)\frac{({\mathbf w}_i-{\mathbf w}_j)\cdot({\mathbf r}_i-{\mathbf r}_j)}{||{\mathbf r}_i-{\mathbf r}_j||^5}\right]
\end{equation}
that are needed to compute the Lyapunov exponent (see below).
As a rule, we use the 4th order integrator with $\Delta t=10^{-2}t_{\rm dyn}$ for models with $N<3\times10^4$ and the 2nd order integrator with $\Delta t=3.3\times10^{-3}t_{\rm dyn}$ for larger system sizes. Throughout this work we assume units such that $G=M=r_c=1$, so that the dynamical time $t_{\rm dyn}=\sqrt{r_c^3/GM}$ and the scale velocity $v_{\rm typ}=r_c/t_{\rm dyn}$ are also equal to 1. Individual particle masses are then $m=1/N$.\\
\indent Typically, in direct $N$-body simulations, the divergence of the Newtonian potential for vanishing interparticle separation, leading to the accumulation of round-off errors in the trajectories, is cured by introducing the softening length $\epsilon_{\rm soft}$ so that the potential at distance $r$ from a particle of mass $m$ becomes $\phi(r)=-Gm/\sqrt{r^2+\epsilon_{\rm soft}^2}$. For our choice of timestep $\Delta t$ we take as optimal value of the softening $\epsilon_{\rm soft}=5\times 10^{-3}r_c$ (for a detailed analysis of the relation between the optimal values of $\Delta t$ and $\epsilon_{\rm soft}$, see \citealt{2011EPJP..126...55D}, and references therein). We verified that, at fixed $\epsilon_{\rm soft}$, the results of the numerical simulations are unchanged when further decreasing $\Delta t$. Tracer particle integrations in frozen potentials are extended up to $t=2000\, t_{\rm dyn}$, while self-consistent equilibrium $N$-body simulations to $t=200\, t_{\rm dyn}$, for all the values of $N$. It is worth noting that the softening of the interactions not only cures the numerical problems related to the divergence of the $1/r$ potential when $r \to 0$, but also makes the system ``less collisional'' than a system with the same $N$ particles but unsoftened interactions, because the strongest collisions where particles become really close to each other have a much smaller impact on the system. However, the two-body relaxation time still grows with $N$ as $t_{2b}\propto N/\ln N$ (see e.g.\ \citealt{2010PhRvL.105u0602G}). Note that, if $\epsilon_{\rm soft}$ is larger than the typical impact parameter of strong encounters (leading to deflections of 90 degrees), then the Coulomb logarithm is no longer proportional to $\ln N$, but becomes the constant $\ln\Lambda =\ln(b_{\rm max}/b_{\rm min})$, where $b_{\rm min}=\epsilon_{\rm soft}$ and $b_{\rm max}$ is typically the scale size of the system, in our case quantified by $r_c$. Note also that the introduction of softening does not reduce the collisional relaxation rate much, precisely because softening only affects the Coulomb logarithm, the latter spanning about two decades upon varying $\epsilon_{\rm soft}$ of six decades in units of the typical scale length of the system (see e.g. \citealt{1990ApJ...349..562H} and \citealt{2001MNRAS.324..273D}).\\
\indent As mentioned in the Introduction, a quantitative measure of the degree of chaotic instability of the dynamics is given by the largest Lyapunov exponent, measuring the exponential growth rate of perturbations of a given trajectory in phase space. 
We compute the numerical estimate of the largest Lyapunov exponent by means of the standard \cite{1976PhRvA..14.2338B} method (see also \citealt{2002ocda.book.....C,2007PhRvL..99m0601G,2013JPhA...46y4005G}) as 
 \begin{equation}\label{lmax}
\lambda_{\rm max}(t) =\frac{1}{L\Delta t}\sum_{k=1}^L\ln\frac{W(k\Delta t)}{W_0}~,
\end{equation}
for a (large) time $t=L\Delta t$, where $W$ is the norm of the $6N$-dimensional vector 
\begin{equation}
\mathbf{W}_{6N}=(\mathbf{w}_i,\dot{\mathbf{w}}_i,...\mathbf{w}_N,\dot{\mathbf{w}}_N),
\end{equation}
for self-consistent simulations, and of the six-dimensional vector
\begin{equation}
\mathbf{W}_6=(\mathbf{w},\dot{\mathbf{w}}),
\end{equation}
for a tracer in a frozen $N$-body model. In both cases, $W_0$ is the value of such norm at $t=0$. The ``true'' largest Lyapunov exponent would correspond to the limit for $L\to\infty$ in Eq.\ \eqref{lmax}, therefore what we compute is, properly speaking, the finite-time Lyapunov exponent $\lambda_{\rm max}(t)$, that may differ from the true asymptotic value\footnote{ If one thinks of the continuum limit for an $N$-body system as its description in terms of one-particle phase-space distribution functions $f$, governed by the CBE, due to the fact that the latter is valid only for $t<t_{2b}$, the limit $L\to\infty$ in Eq.\ (\ref{lmax}) becomes questionable; $t_{2b} \to \infty$ too, but the limits $N\to\infty$ (continuum) and $L\to\infty$ may not commute.} $\lambda_{\rm max} = \lim_{t \to \infty}\lambda_{\rm max}(t)$. However, we checked that $L$ is large enough for the result to appear relaxed to its asymptotic value, so that we may be confident that our long-time results are a good approximation to $\lambda_{\rm max}$.  Hereafter, to avoid confusion, we will use $\lambda_{\rm max}$ to denote the largest Lyapunov exponent of a single particle orbit in
\begin{figure*}
\includegraphics[width=0.96\textwidth]{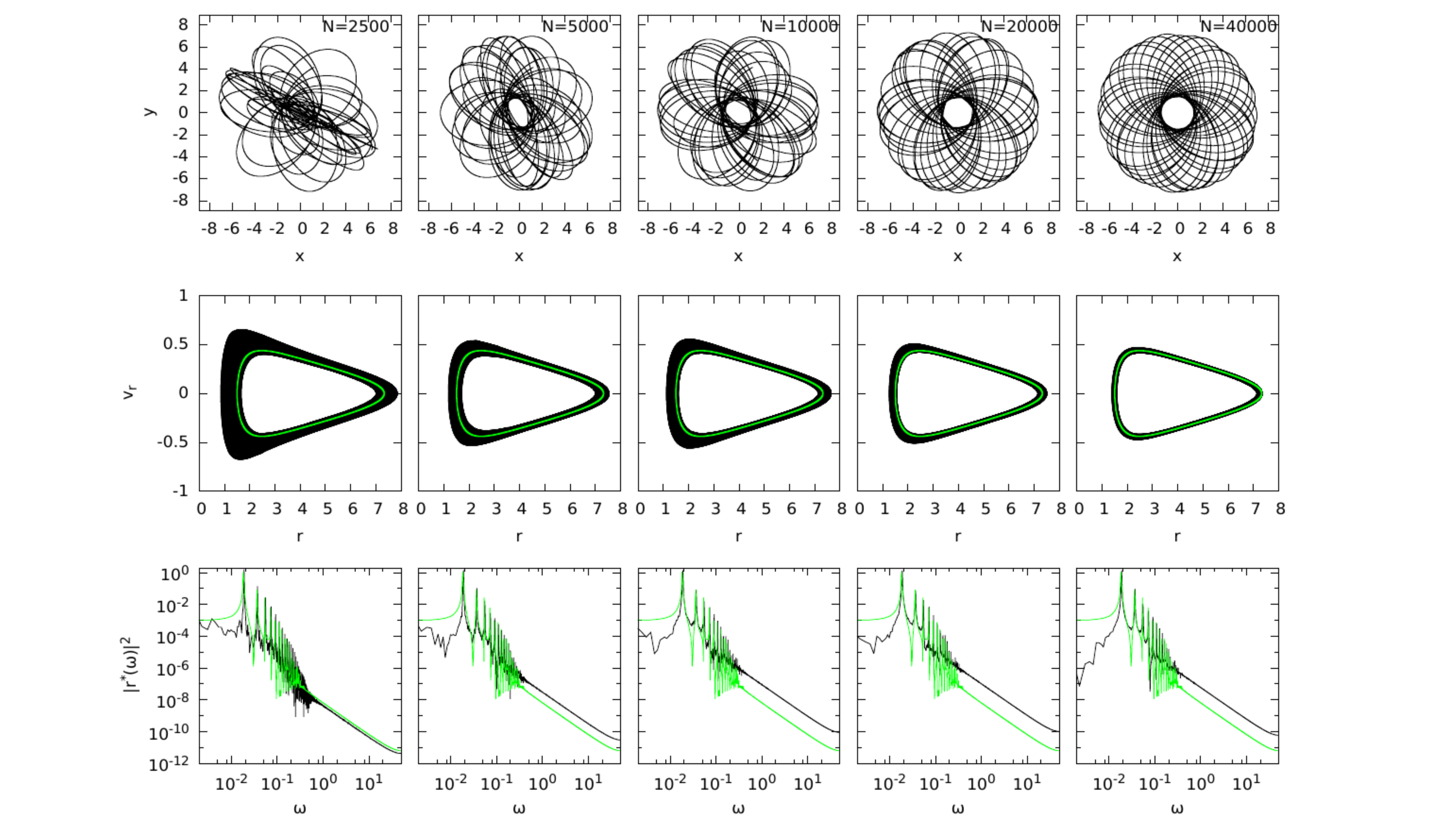}
\caption{Top row: projections on the $x,y$ plane of five tracer orbits starting with the same initial condition $(\mathbf{r}_0,\mathbf{v}_0)$, with energy per unit mass $\mathcal{E}=-0.1$, evolved in a frozen $N$-body potential generated by a discrete Plummer distribution of $N=2.5\times10^3$, $5\times10^3$, $\times10^4$ $2\times10^4$ and $4\times10^4$ particles. Middle row: orbit section in the $r,v_{r}$ plane. Bottom row: squared modulus of the Fourier spectrum of the radial coordinate $r$. In all plots the black lines refer to the discrete models, while the green (gray) ones to the same initial condition propagated in the parent smooth potential.}
\label{figplumtrac}
\end{figure*}
\begin{figure*}
\includegraphics[width=0.96\textwidth]{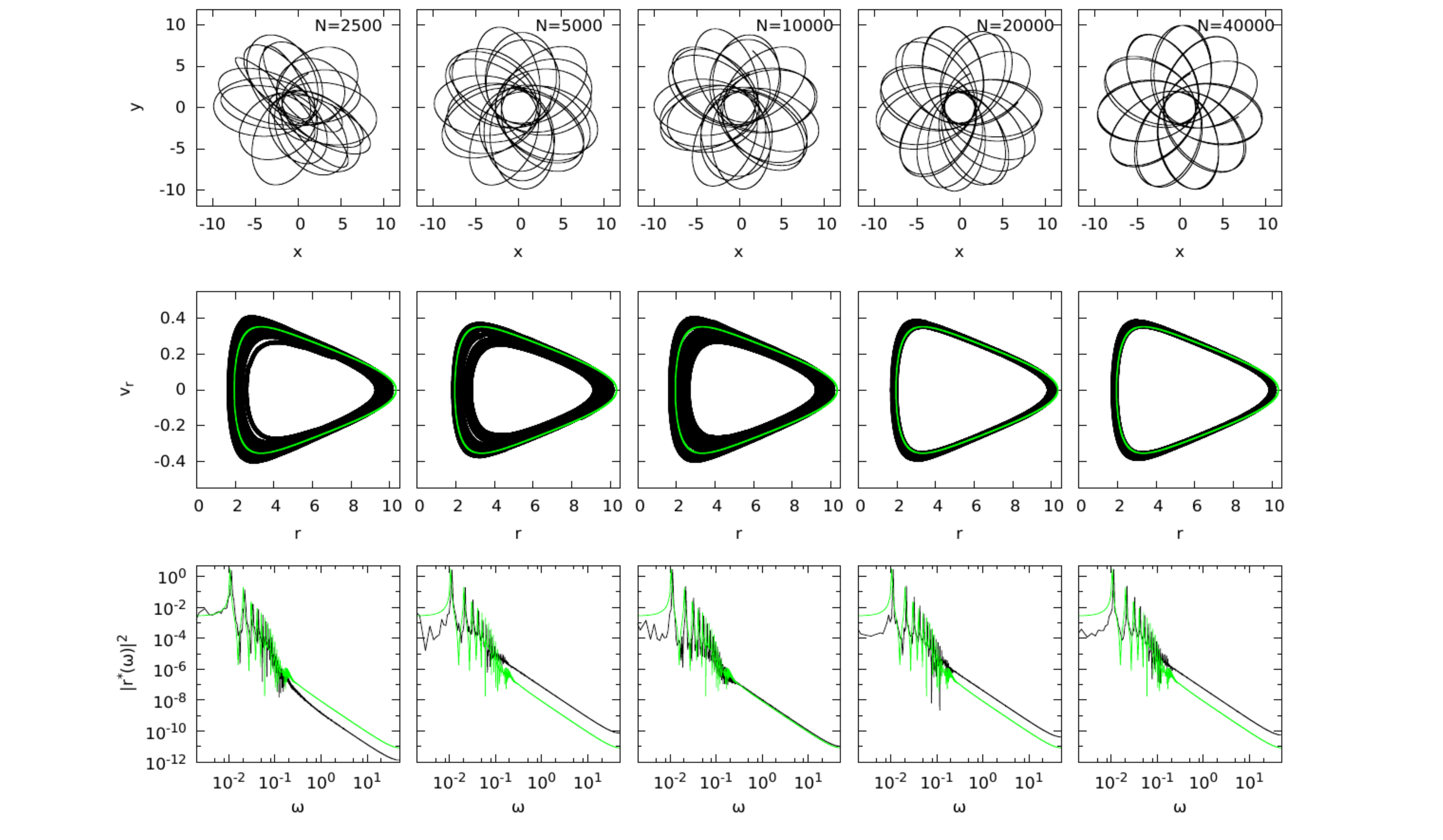}
\caption{Same as in Fig.\ \ref{figherntrac} but for a Hernquist model.}
\label{figherntrac}
\end{figure*}
 a frozen or smooth potential, and $\Lambda_{\rm max}$ for the largest Lyapunov exponent of the full $6N$-dimensional self-consistent problem.
To improve convergence, following \cite{1976PhRvA..14.2338B}, the vector $\mathbf{W}_{6N,6}$ is periodically renormalized to $W_0$. In all simulations presented here, the renormalization procedure is done every $10\Delta t$. The final value attained by $\lambda_{\rm max}$ (or $\Lambda_{\rm max}$) is independent of the frequency of this operation and the value of $W_0$, that we 
fix to unity in all simulations shown here.\\
\indent We note that in previous works several authors (see e.g.\ \citealt{2004CeMDA..90..147S,2004PhRvS...7a4202K,meschiari06}) used, in order to evaluate Eq.\ (\ref{lmax}) for a tracer particle, the difference between two realizations of the same orbit starting with initial conditions $(\mathbf{r}_0,{\mathbf{v}}_0)$ and $(\mathbf{r}^\prime_0,{\mathbf{v}}^\prime_0)$, i.e. $\tilde{\mathbf{W}}_6=(\mathbf{r}-\mathbf{r}^\prime,{\mathbf{v}}-{\mathbf{v}}^\prime)$ instead of the tangent vectors. By doing so, the value of the largest Lyapunov exponent $\tilde{\lambda}_{\rm max}$ strongly depends on the choice of $\tilde{W}_0$, at variance with its counterpart evaluated using the tangent dynamics that, in general, has a different value and is independent of $W_0$ (see e.g.\  \citealt{MEI2018108}). In order to assess the magnitude of such discrepancy, we have performed some test simulations where two realizations of a given trajectory with different initial distance $\tilde{W}_0$ in the range $(10^{-13},1)$ were integrated in the frozen potential generated by a distribution of $N$ particles extracted from a Plummer distribution, for various $N$. We computed the largest Lyapunov exponent over $2000\, t_{\rm dyn}$ by means of the standard expression (\ref{lmax}) using the tangent vectors, as well as using the difference between the two realizations, having set in all cases $\mathbf{W}_6=\tilde{\mathbf{W}}_6$ at $t=0$. We observed that, as expected, the values attained by the Lyapunov exponent computed in the ``correct'' way is independent on the initial value of the norm for all orbit energies and system sizes, as shown in the left panel of Fig.\ (\ref{test1}) for two particles with initial energies $\mathcal{E}=-0.1$ and $-0.5$ in a frozen Plummer model with $N=10^5$. On the contrary, when evaluating $\tilde{\lambda}_{\rm max}$ using the difference between two initially close realizations of the same orbit, its value is significantly different from $\lambda_{\rm max}(t)$ at all times for $\tilde{W_0}>10^{-13}$. This is exemplified in the right panel of Fig.\ (\ref{test1}), where $\lambda_{\rm max}(t)-\tilde{\lambda}_{\rm max}(t)$ is plotted as function of time, and it can be clearly seen to converge to zero only for the smallest choice of $\tilde{W}_0$. 
\begin{figure*}
\includegraphics[width=0.97\textwidth]{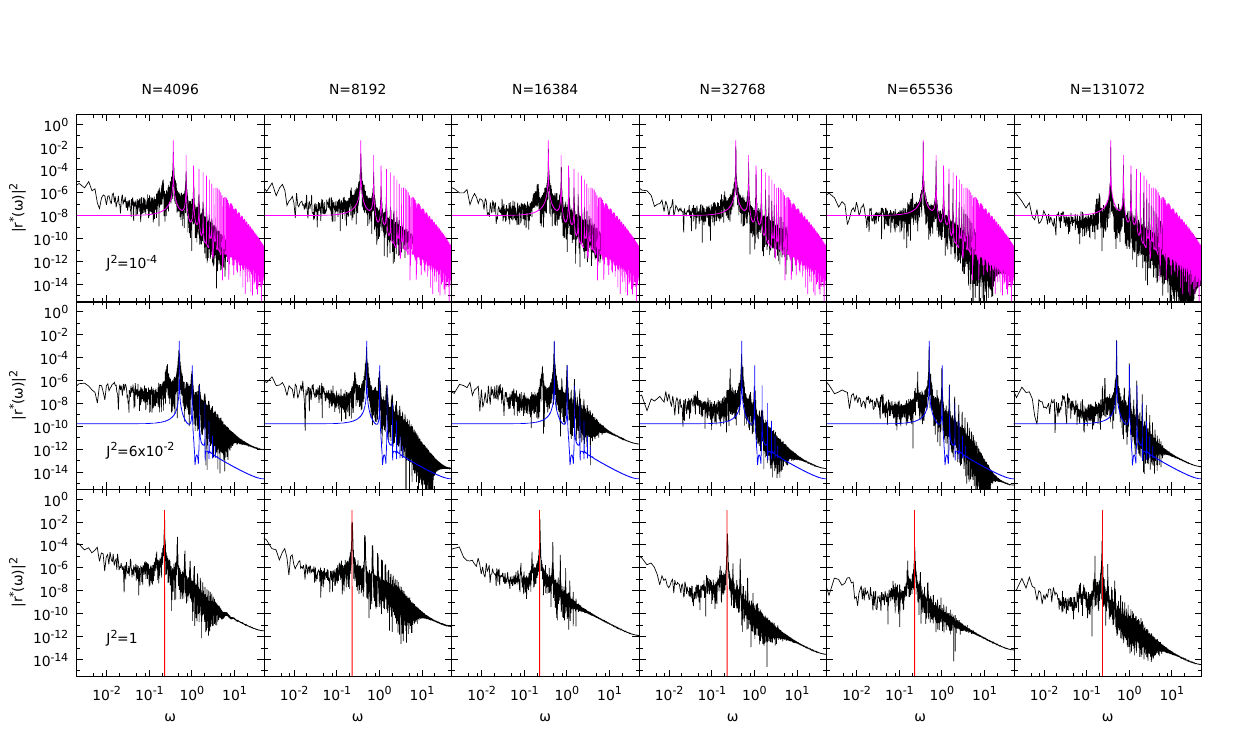}
\caption{Fourier spectra of radial coordinates of orbits with $\mathcal{E}=-0.8$ propagated in a frozen Plummer model with increasing $N$ and different values of angular momentum $J^2=10^{-4}$ (upper panels), $6\times 10^{-2}$ (middle panels), and $1$ (bottom panels). The black curves refer to the orbit in the frozen $N$-body potential, while the coloured (gray) ones to the corresponding orbit propagated in the smooth Plummer potential.}
\label{spettri}
\end{figure*}
\section{Simulations and results}
\label{sec:simulations}
\subsection{Orbits in frozen $N$-body potentials}
Single particle orbits in frozen $N$-body self-gravitating systems have been extensively used to study the dynamics in triaxial systems for which analytic formulations of the phase-space distribution function are unknown (see e.g.\ \citealt{2004PhRvS...7a4202K,2013MNRAS.436.1201C,2014MNRAS.438.2201M,2016MNRAS.458.3578M,2017ApJ...850..145C,2018A&A...612A.114P}, and references therein) and are usually constructed numerically using orbit libraries
\begin{figure*}
\includegraphics[width=0.97\textwidth]{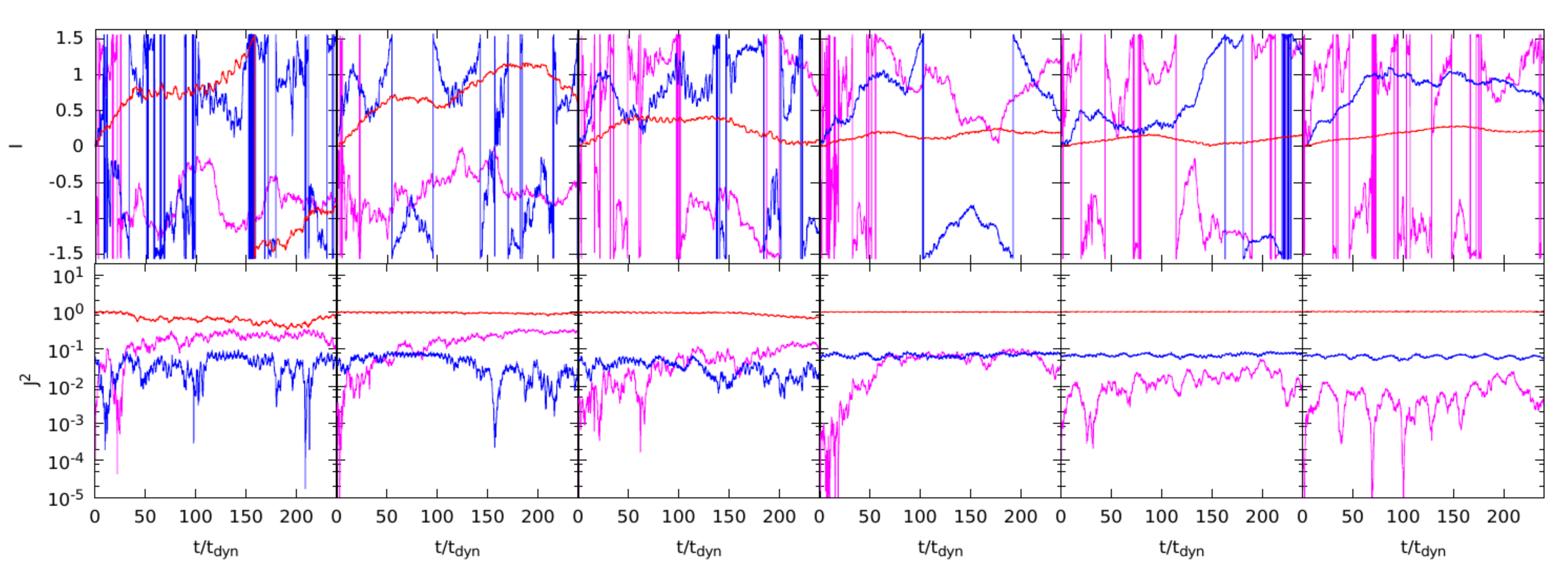}
\caption{Evolution of the orbital inclination $I$ (top) and squared angular momentum $J^2$ (bottom) for the same orbits as in Fig. \ref{spettri}.}
\label{ij2}
\end{figure*}
 (\citealt{1979ApJ...232..236S,2002PhDT.........8T,2003astro.ph..5005T}). It is well known (see e.g.\ \citealt{1996ApJ...471...82M,1998NYASA.858...48M,1998ApJ...506..686V}) that triaxial potentials are associated to mixed phase-space, admitting regions with both regular and chaotic orbits. When it comes to the study of discrete $N$-body models, it is not always easy to determine on which extent the chaos is 
due to the discreteness or to the finite deviation from spherical symmetry.\\
\indent In this paper we study the orbital structure and the scaling of the the largest Lyapunov exponent as a function of the number of particles $N$ in \textit{spherical} models, starting with their frozen realizations.
 We have evolved the same initial condition $(\mathbf{r}_0,\dot{\mathbf{r}}_0)$ in the frozen $N$-body realizations of Plummer and Hernquist density profiles for $10^2\leq N\leq 3\times 10^7$.  In Figs.\ \ref{figplumtrac}-\ref{figherntrac} we show, for Plummer and Hernquist models, respectively, the same initial conditions propagated in the potential of a frozen $N$-body model for $N=2.5\times 10^3$, $5\times 10^3$, $10^4$, $2\times 10^4$ and $4\times 10^4$. For both models, as $N$ increases the orbit projection in the $x,y$ plane (upper rows) becomes more and more regular and markedly centrophobic (cfr.\ analogous plots in \citealt{2001PhRvE..64e6209K,2002PhRvE..65f6203S,2004PhRvS...7a4202K}). Consistently, and as expected, the radial phase-space sections $r,v_r$ (middle rows) show that the orbit propagated in the discrete frozen potential (black dots) approaches that in the smooth potential (green/gray dots) for increasing values of $N$. As a general trend, and for both choices of $\rho(r)$ the phase-space sections show a larger diffusion from the ``analytic orbit'' at low radii and are bound 
\begin{figure}
\includegraphics[width=0.98\columnwidth]{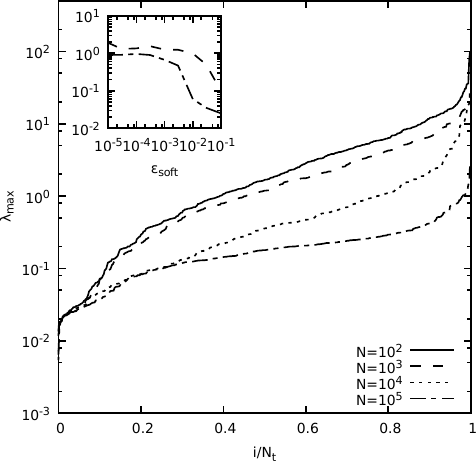}
\caption{Ordered plot of the largest Lyapunov exponents for a set of $N_t=500$ tracers propagating in frozen $N$-body Plummer potential with $N=10^2$, $10^3$, $10^4$ and $10^5$ (main plot). Maximal Lyapunov exponent for an orbit with $\mathcal{E}=-1$ as function of the softening length $\epsilon_{\rm soft}$ for the $N=10^3$ and $10^5$ casesd (inset).}
\label{figord}
\end{figure}
between two limit curves up to $t=2\times 10^3t_{\rm dyn}$ (the time to which the frozen $N$-body integrations are extended). On the other hand, the behaviour of Fourier spectra of the radial coordinate $r$, $|r^*(\omega)|^2$ (bottom rows) has a less trivial trend with the system size $N$. For both Plummer and Hernquist density profiles, the peak in the spectrum corresponding to the fundamental radial frequency $\omega_{r}$ does not appear to be significantly shifted with respect to that of the orbit in the smooth potential. The low and high frequency tails do not appear to have any significant trend with $N$ for the Plummer model, while appear to reproduce better the spectrum of the orbit integrated in the smooth potential for intermediate $N$ (e.g., $N=10^4$ in Fig.\ \ref{figherntrac}). For different values of the integration parameters, such as the order of the symplectic integrator, the timestep $\Delta t$ and the softening length $\epsilon_{\rm soft}$, we observe a qualitatively similar behviour, with a net tendency to have a slight shift in the fundamental orbital frequencies for increasing values of $\epsilon_{\rm soft}$.\\
\indent In general, the mismatch between the Fourier spectra of orbits propagated in analytic smooth potentials and frozen $N$-body space does not decrease for increasing $N$ (at least for the system sizes considered here in the range $10^2 \div 10^7$), but for fixed initial orbit energy (per unit mass) $\mathcal{E}$, orbits starting with higher values of the initial angular momentum $J$ seem to have power spectra more similar to those of the parent analytical orbit. This is represented in Figure \ref{spettri} where, as an example, we show $|r^*(\omega)|^2$ for three different initial conditions evolved in a frozen Plummer model. In all cases the orbits have comparable values of energy per unit mass, while the initial angular momentum\footnote{Note that $\mathbf{J}$ is conserved only for orbits propagated in the smooth potential.} $\mathbf{J}_0$ has significantly different values in the three rows of curves.
\begin{figure}
\includegraphics[width=\columnwidth]{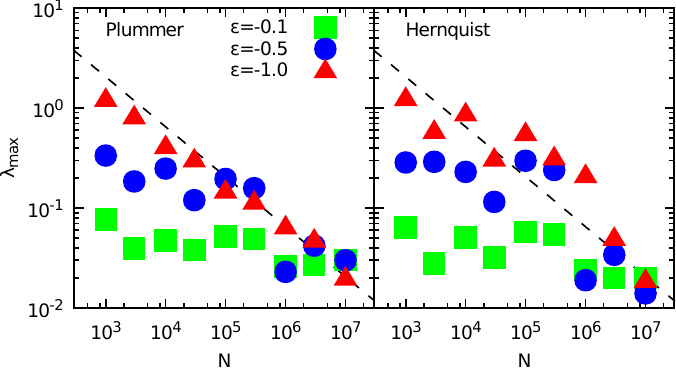}
\caption{Largest Lyapunov exponent $\lambda_{\rm max}$  for a tracer orbit with ${\mathcal{E}=-0.1,-0.5}$ and $-1$ in a frozen $N$-body model as a function of $N$, for Plummer (left panel) and Hernquist (right panel) profiles. The dashed line is the $N^{-1/2}$ law.}
\label{figlmaxtrac}
\end{figure} 
 It is evident that for tracer orbits starting with vanishing initial angular momentum (upper row), and the maximum attainable angular momentum (bottom row), increasing the value of $N$ has little effect on improving the matching between the discrete Fourier spectra with that of the orbit in the smooth potential. Only the cases with $N\geq 32768$ and $J_0=1$ stand out for not presenting the higher frequency peaks appearing for lower values of $N$. For completeness, in Fig. \ref{ij2} we present for the same systems of Fig. \ref{spettri} the evolution of their orbital inclination $I$ (i.e. the angle with respect to the reference plane $z=0$) and squared angular momentum $J^2$. It appears clearly that for low angular momentum orbit, the orbital plane undergoes wild oscillations that do not appear to damp out for increasing $N$. The angular momentum modulus itself varies strongly even on time scles as short as $50t_{\rm dyn}$. For orbits starting with large values of $J_0$ (closer and closer to the circular orbit with $v_{\rm circ}=\sqrt{rGM(r)}$), the conservation of $J^2$ improves for increasing $N$ but the inclination $I$ still presents appreciable changes, reason for which there is still considerable ``noise'' in the power spectra of $r$ for large values of $N$ in Figs. \ref{figplumtrac}-\ref{spettri}. Tuning the force softening $\epsilon_{\rm soft}$ at fixed $N$ and tracer orbit initial parameters reflects on the structure 
\begin{figure*}
\includegraphics[width=0.95\textwidth]{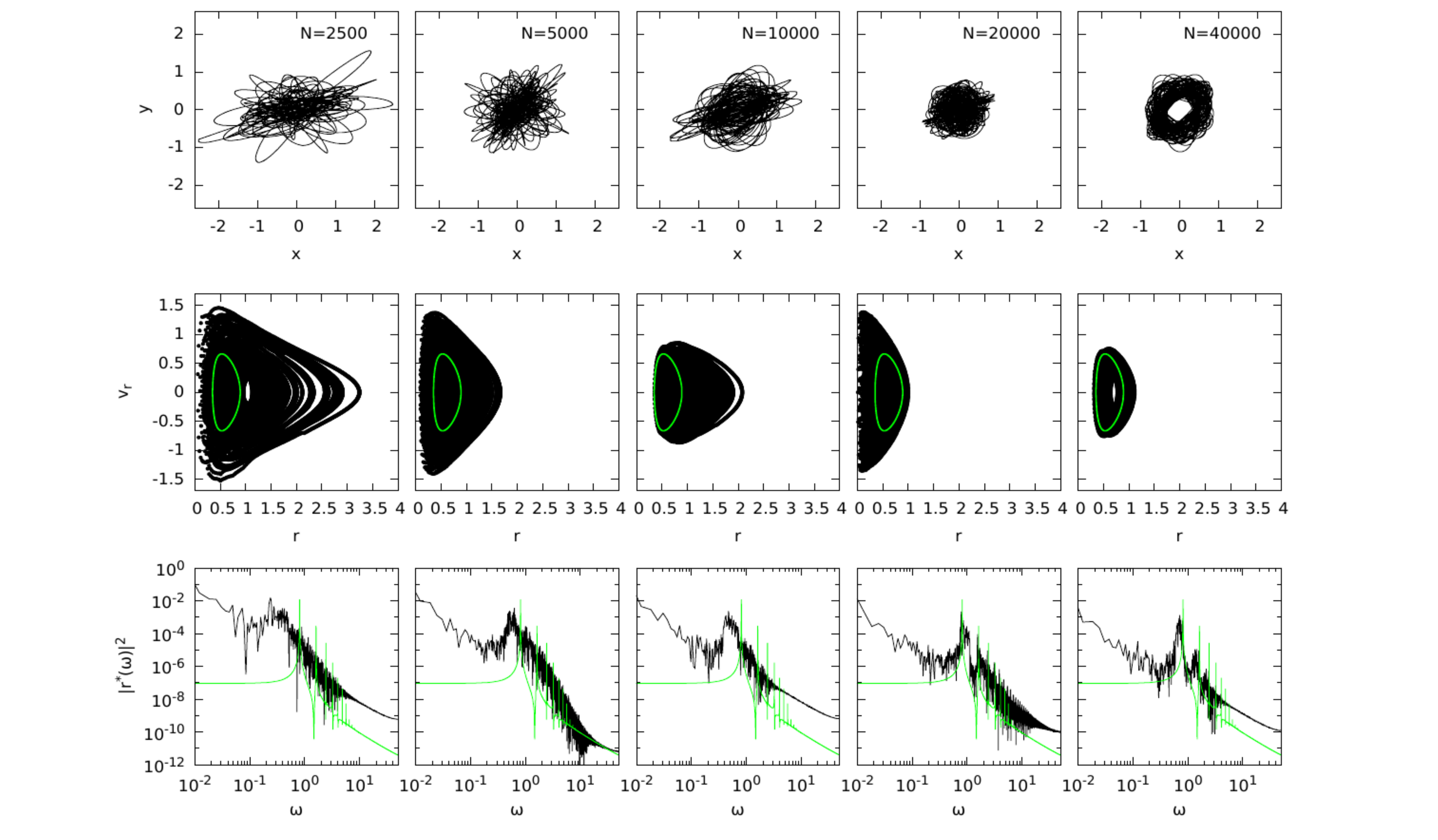}
\caption{Top row: projections on the $(x,y)$ plane of five tracer orbits starting with the same initial condition $(\mathbf{r}_0,\mathbf{v}_0)$, again with initial energy per unit mass $\mathcal{E}=-0.1$ in a self-consistent $N$-body simulation with initial conditions sampled from an isotropic Plummer model with $N=2.5\times10^3$, $5\times10^3$, $\times10^4$ $2\times10^4$ and $4\times10^4$ particles. Middle row: orbit section in the $r,v_{r}$ plane. Bottom row: squared modulus of the Fourier spectrum of the radial coordinate $r$.}
\label{figplumself}
\end{figure*}
\begin{figure*}
\includegraphics[width=0.95\textwidth]{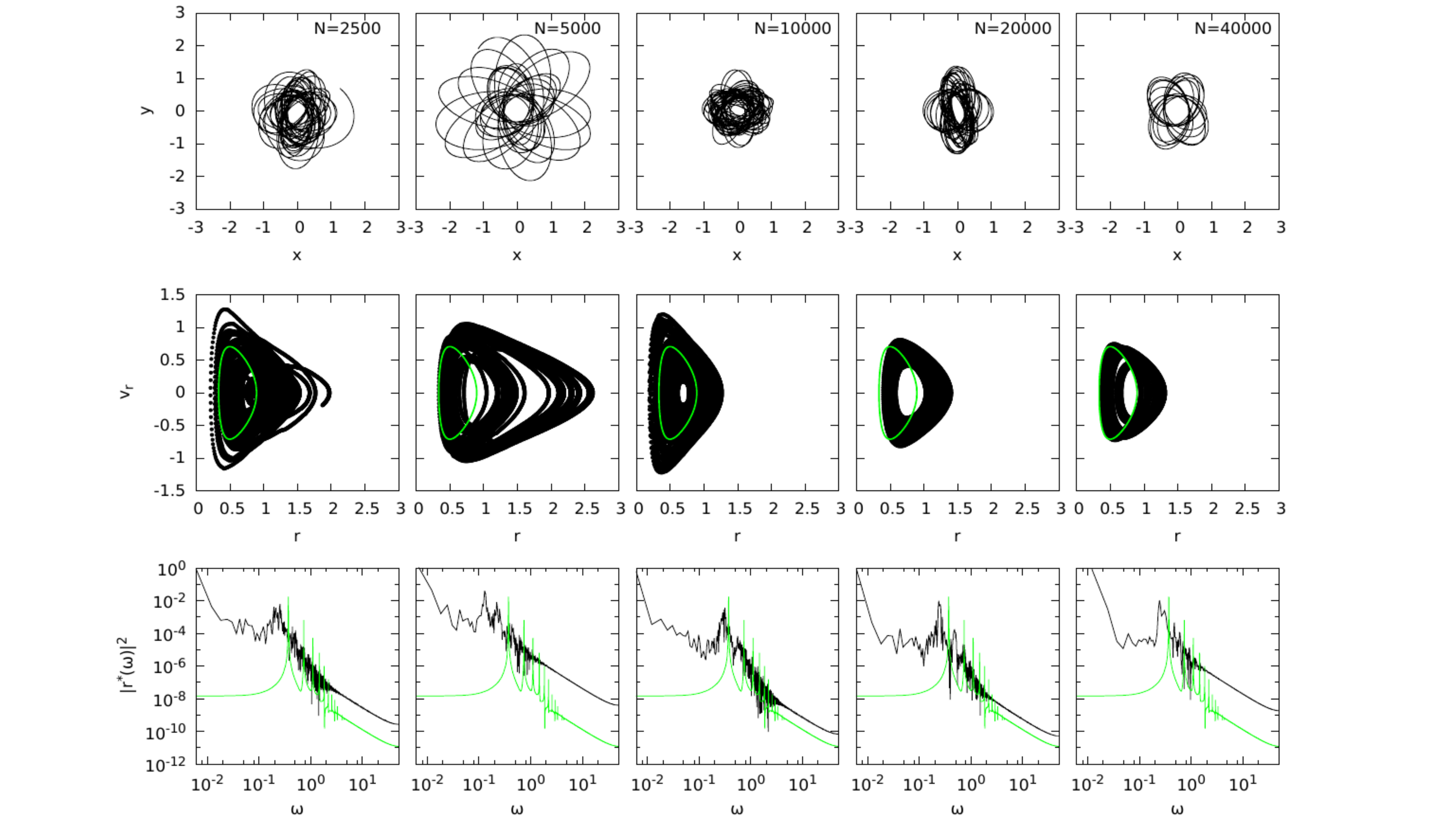}
\caption{The same as in Fig.\ \ref{fighernself} for a Hernquist model.}
\label{fighernself}
\end{figure*}
 of the power spectra of the radial coordinate $r$ in a shift of the fundamental frequency to slightly lower values of $\omega$ and a narrowing of the associated peak for $\epsilon_{\rm soft}$ increasing towards $0.1r_c$ and decreasing towards $10^{-6}r_c$ (the limit values of $\epsilon_{\rm soft}$ used in our numerical simulations). For intermediate values (say in the range $3\times 10^{-4}r_c<\epsilon_{\rm soft}<3\times 10^{-2}r_c$) we do not observe any appreciable variation in the discrepancies with the spectra of the analytical orbit.\\
\indent For both frozen Plummer and Hernquist models we have evaluated the maximal Lyapunov exponent $\lambda_{\rm max}$ for orbits with different values of the energy per unit mass $\mathcal{E}$ and angular momentum $\mathbf{J}$. As it is evident from Figure \ref{figord}, where we show the ordered plot of the maximal Lyapunov exponents for $N_t=500$ orbits propagated for $10^3t_{\rm dyn}$ in frozen Plummer potentials with different values of $N$, where with increasing system size, the distribution of $\lambda_{\rm max}$ of the tracer particles attains systematically lower values. This is particularly evident for the larger values of the Lyapunov exponents, usually associated with smaller values of the orbital energy per unit mass $\mathcal{E}$.\\
\indent In order to quantify such behaviour as function of $N$ and $\mathcal{E}$, in Figure \ref{figlmaxtrac} we show the trend with $N$ of $\lambda_{\rm max}$ for three typical orbital energy values $\mathcal{E}=-0.1,$ $-0.5$, $-1$ obtained averaging over 100 independent realizations. The value of the Lyapunov exponent seems to be almost independent of the number of particles $N$ for weakly bound orbit (i.e., $\mathcal{E}=-0.1$ in this case, see also Fig. \ref{figord} for $i/N_t\to 0$), while it seems to approach a $\lambda\propto N^{-1/2}$ decay as the tracer orbit is more and more bound, independently of the specific density profile at hand. Moreover, we observe that decreasing the softening length at fixed $N$ is always associated with a systematic increase of the values of $\lambda_{\rm max}$ for the tracer orbit. However, for $\epsilon_{\rm soft}\lesssim l_{\rm int}$ (the mean inter-particle distance within $r_c$), the values of the Lyapunov exponents appear to accumulate to a limit value, rather than increasing indefinitely (see inset in Fig. \ref{figord}). 
\subsection{Self-consistent equilibrium models} 
\begin{figure}
\includegraphics[width=\columnwidth]{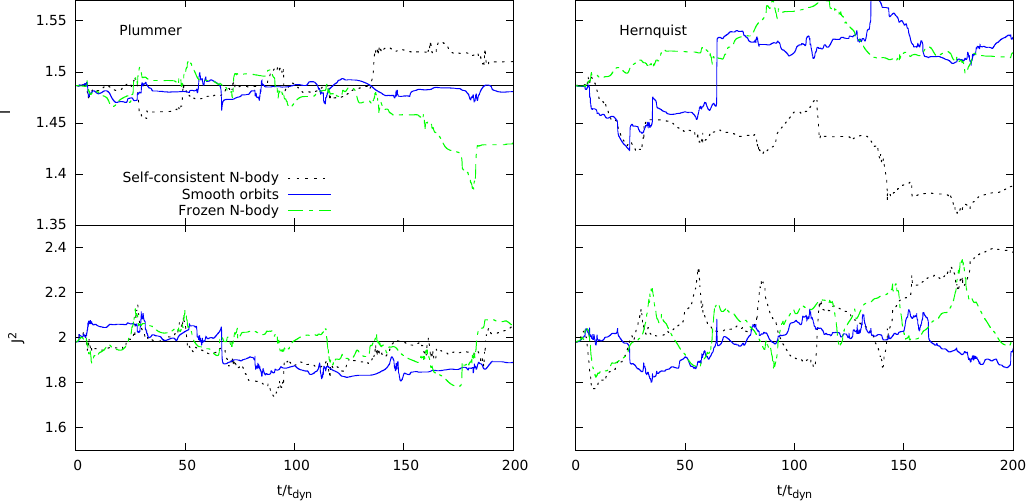}
\caption{Time evolution of the orbital inclination $I$ (upper panels) and the square of the angular momentum $J^2$ (lower panels) for a tracer particle in a frozen $N$-body potential (dotted-dashed lines), a self consistent model (dotted line) and a system of non-interacting orbits moving in a smooth potential (solid  lines), for Plummer (left) and Hernquist (right) density profiles. In all cases $N=16348$}
\label{figj2}
\end{figure}

\begin{figure*}
\includegraphics[width=0.95\textwidth]{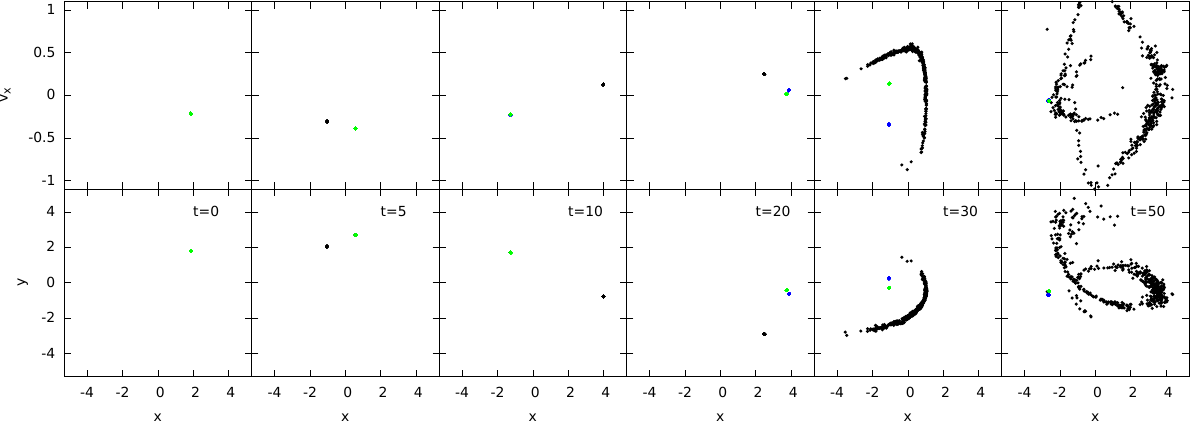}
\caption{Projections in the $x,v_x$ (top row) and $x,y$ planes of a distribution non-interacting tracers propagated in a self-consistent Plummer model (black points), a frozen Plummer model (green/light gray points) and a system of non interacting orbits in a smooth Plummer potential (blue/dark gray points), at (from left to right) $t=0$, 5 10, 20, 30 and 50. In all cases $N=20000$ and $N_t=500$.}
\label{diffusion}
\end{figure*}

\begin{figure*}
\includegraphics[width=0.95\textwidth]{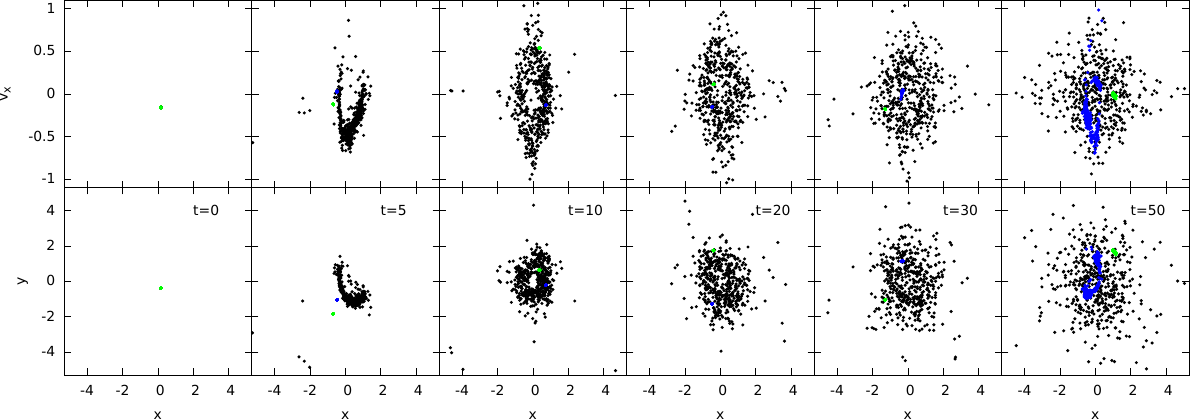}
\caption{Same as Fig. \ref{diffusion} but for a Hernquist model.}
\label{diffusion2}
\end{figure*}

\begin{figure}
\includegraphics[width=\columnwidth]{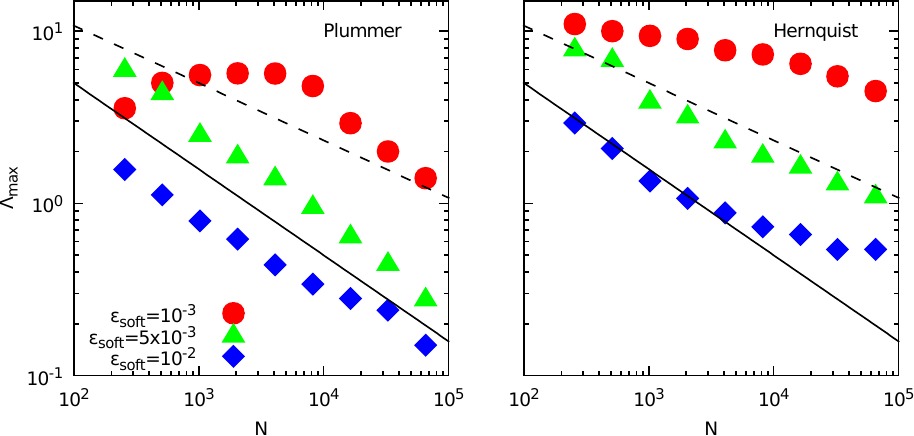}
\caption{Largest Lyapunov exponent $\Lambda_{\rm max}$ for an isotropic self-consistent $N$-body model as function of $N$ and different values of the force softening $\epsilon_{\rm soft}$, for Plummer (left panel) and Hernquist (right) density profiles. The solid and dashed lines mark the theoretical $N^{-1/2}$ and $N^{-1/3}$ laws, respectively.}
\label{figlmaxself}
\end{figure}

We repeated the analysis described above for active (i.e., their contribution to the force field is accounted) and non-interacting tracer particles with the same initial conditions $(\mathbf{r}_0,{\mathbf{v}}_0)$ in self-consistent $N$-body runs with the same values of $N$ (see Figs.\ \ref{figplumself}, \ref{fighernself}). We find that, starting from the same initial conditions, an active particle and a tracer particle are virtually indistinguishable for $N>2000$.\\
\indent  From the point of view of the orbital structure, for both Plummer and Hernquist models, the situation in this case is more consistent with what one would expect, with more similar spectra at larger $N$. In particular, as $N$ increases individual particle orbits become more regular (even though less rapidly than in frozen $N$-body models) and explore less and less phase space.\\
\indent Additionally, we have computed the maximal Lyapunov exponent $\lambda_{\rm max}$ for different tracer particles moving in an active self-consistent model and its frozen $N$-body and smooth orbits counterparts, for both choices of the density profile and different values of energy per unit mass $\mathcal{E}$ and angular momentum $\mathbf{J}$.\\ 
\indent The distributions of Lyapunov exponents as a function of the initial particle energy and angular momentum, for fixed profile and system size $N$, are remarkably similar for 
\begin{figure}
\includegraphics[width=0.85\columnwidth]{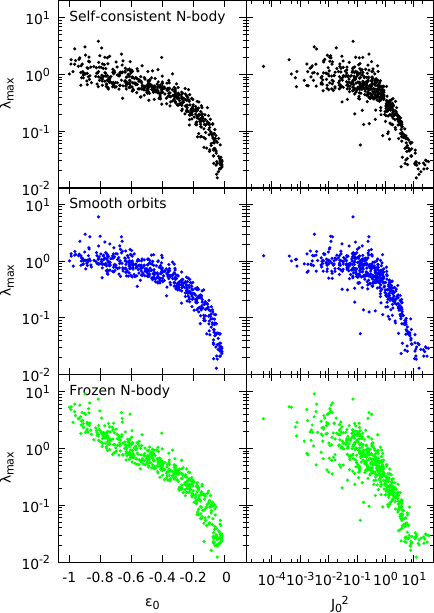}
\caption{Largest Lyapunov exponent $\lambda_{\rm max}$  for a system of 500 tracer particles moving in a  isotropic self-consistent $N$-body model (top panels), in a system of non-interacting particles propagated in a smooth potential (middle panels), and in a frozen $N$-body model (bottom panels), as a function of the initial energy per unit mass $\mathcal{E}_0$, (left column) and initial value of the squared angular momentum $J^2_0$ (right column). In all cases $N=10000$ and Plummer density profile.}
\label{snapshot}
\end{figure}
\begin{figure}
\includegraphics[width=0.96\columnwidth]{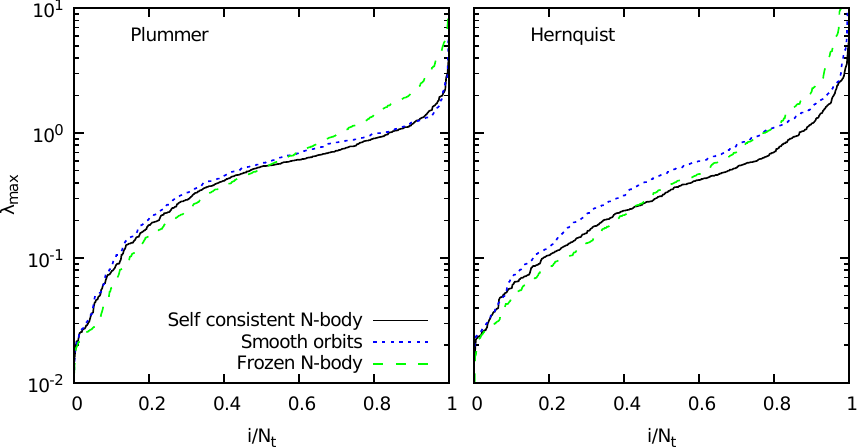}
\caption{Ordered plot of the largest Lyapunov exponents for a set of $N_t=500$ tracers propagating in active $N$-body models (solid lines), non-interacting systems of smooth orbits (dotted lines) and frozen $N$-body potentials (dashed lines), for $N=10000$ and Plummer (left) and Hernquist (right) density profiles.}
\label{ord2}
\end{figure}
tracers propagated in self-consistent $N$-body and smooth orbit systems, with the frozen $N$-body case showing instead a different slope in the $\lambda_{\rm max},\mathcal{E}_0$ and $\lambda_{\rm max},J^2_0$ planes, with larger exponents associated with smaller initial energies as shown in Fig. \ref{snapshot} for the $N=10000$ Plummer case. This reflects also in the structure of the ordered plot of $\lambda_{\rm max}$, as presented in Fig.\ \ref{ord2} for the Plummer (left) and Hernquist (right) density profiles and $N=10000$ (i.e., the self-consistent 
\begin{figure}
\includegraphics[width=\columnwidth]{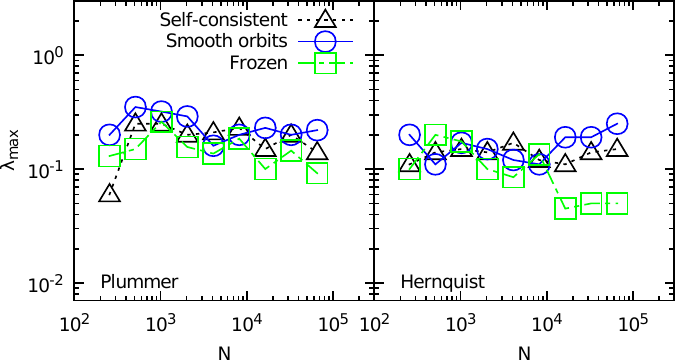}
\caption{Largest Lyapunov exponent $\lambda_{\rm max}$  for a tracer particle moving in an isotropic self-consistent $N$-body model (triangles), in a system of non-interacting particles propagated in a smooth potential (circles), and in a frozen $N$-body model (squares), as a function of $N$, for Plummer (left panel) and Hernquist (right panel) density profiles.}
\label{figlmaxsingle}
\end{figure}
and smooth orbit cases have the same slopes). In general, we observe again that for weakly bound particles with large angular momentum, the values themselves attained by $\lambda_{\rm max}$ for given $N$ do not present substantial difference from one model to one another, as shown in Fig. \ref{figlmaxsingle} for the $\mathcal{E}\simeq -0.3$ and $J^2\simeq 2$ case. For $N>10^4$ such orbits do not appear to have significant changes in both the magnitude and the orientation of $\mathbf{J}$, as shown in Fig.\ \ref{figj2}. Vice-versa, the value of $\lambda_{\rm max}$ for tracer particles in self-consistent systems becomes more strongly dependent on the initial energy $\mathcal{E}_0$ for more bound particles, the latter having in general more chaotic trajectories.\\ 
\indent We interpret all these facts as a hints that the continuum limit shall be questioned at least concerning the concept of $N$-body chaos, in particular, with respect to the usefulness of frozen $N$-body models as surrogates of an equilibrium finite-$N$ body problem. The phase-space transport\footnote{ Collisional relaxation is associated with diffusive $t^{1/2}$ growth in action-angle space, collisionless relaxation (i.e., phase mixing) to a linear growth of phases.} of initially localized orbit families should be quite different in the two cases since in a frozen model the single particle energy is always conserved independently 
of $N$, while in active $N$-body models particles could exchange energy, even in the limit  $N\to\infty$. This is exemplified in Figs.\ \ref{diffusion}-\ref{diffusion2} (for Plummer and Hernquist profiles, respectively) where we show at different times ($t=0$, $5$, $10$, 20, 30 and $50 \, t_{\rm dyn}$) the positions in the $(x,y)$ and $(x,v_x)$ planes of an initially localized bunch of $N_t=500$ tracers propagated in a $N=20000$ self consistent model (black points), a system of non-interacting particles in a static smooth potential (blue/dark gray points) and a frozen (green/light gray points). Tracer particles in self-consistent simulations rapidly explore a large portion of configuration and phase-space in the Hernquist case (Fig.\ \ref{diffusion2}), while they start spreading at later stages, say at around $30t_{\rm dyn}$ in the Plummer case (Fig.\ \ref{diffusion}) for comparable values of initial energies and angular momentum. Tracer particles moving in frozen $N$-body systems or interacting with moving particles in a fixed smooth potential, tend instead to remain clustered for long times in both configuration and phase-spaces,
 with only the case of non-interacting orbits in smooth Hernquist potential showing some considerable spread at $t=50t_{\rm dyn}$. As a matter of fact, this is because self-consistent models do have collective effects, that may also be enhanced by the discreteness-induced noise (\citealt{1998MNRAS.297..101W,2016MNRAS.461.1745E}), while frozen $N$-body models have nothing but intrinsic discreteness noise. The third case, represented by non-interacting orbits moving in a fixed smooth potential, lies in between as its effective field on a tracer particle is time dependent. In order to check whether the spread in phase-space over the integration times is consistent with a diffusive process (associated with two body collisions), we have computed the evolution of the tracers' mean velocity $\bar{v}$ and root-mean squared velocity $v_{\rm rms}\equiv\sqrt{\sum_{i=1,N_t}(v_i-\bar{v})^2}/N$, as well as their mean energy and angular momentum $\bar{\mathcal{E}}$ and $\bar{J}$. We find that, independently of the density profile, $\bar{v}$ has an oscillatory behaviour while $v_{\rm rms}$ grows logarithmically up to $\sim 20t_{\rm dyn}$, saturating for the frozen $N$-body case and rapidly increasing with a steep power-law trend for the self-consistent and smooth orbit cases. In general, the evolution of the tracer r.m.s. velocity is incompatible with a growth $\propto t^{1/2}$ as expected from a diffusive process \`a la Chandrasekhar. The evolution of the average CBE invariants, by definition not subjected to phase-mixing at variance with particle velocities, has an oscillatory behaviour in frozen systems independently of $N$, while it is compatible with a power-law for the smooth potential and the self consistent models with $N\lesssim 1000$.\\
\indent We argue that the discrepancies between analytical orbits in smooth potentials and tracer orbits in frozen and live $N$-body models are essentially an effect of the overall discreteness and chaoticity of the latter potentials, rather than an effective collisional evolution, in direct $N$-body simulations, (see also \citealt{1997ApJ...480..155H}).\\
\indent Let us now consider the largest Lyapunov exponent $\Lambda_{\rm max}$ for an active self-consistent model (Fig.\ \ref{figlmaxself}), and especially its dependence on $N$ and on the nature of the density profile.
 The degree of chaoticity of such a system strongly depends on the nature of the density profile, with cuspy models being 
in general more chaotic for sufficiently large $N$.\\ 
\indent In all cases the largest Lyapunov exponent decreases with $N$ (at least for $N>2\times 10^3$) resulting in increasing values of the Lyapunov time (that is, $\Lambda^{-1}_{\rm max}$), that remains within one order of magnitude around the dynamical time. In order to check the whether such behaviour is affected by the short distance regularization of the $1/r^2$ force law, we have performed simulations with same initial 
conditions and different values of the softening length $5\times 10^{-4}\leq \epsilon_{\rm soft}\leq 3\times10^{-2}$ 
(and associated optimal timestep, see \citealt{2011EPJP..126...55D}). As expected, larger values of $\epsilon_{\rm soft}$ results in lower values of $\Lambda_{\rm max}$ for fixed density profile and particle number $N$. Curiously, for the Plummer model (left pannel in Fig.\ \ref{figlmaxself}) we observe a slight {\it increase} of $\Lambda_{\rm max}$ with $N$ for the smaller value of $\epsilon_{\rm soft}\approx10^{-3}$ for $N<2\times 10^3$. We verified that such increase holds for even smaller values of the softening length (yielding remarkably similar values of $\Lambda_{\rm max}$, not pictured here), however what happens in this case for larger values of $N$ is left undetermined, due to the prohibitively small values of the associated optimal timestep $\Delta t$ to be used for the simulations. We conjecture that this latter issue might have led to speculate that $\Lambda_{\rm max}$, and more in general the Lyapunov exponent associated to single particle trajectories are indeed constant or slightly increasing functions of $N$. Remarkably, for sufficiently small values of the softening length we find evidence of a crossover between a Lyapunov exponent nearly constant with $N$ at small $N$'s and a decreasing one at larger $N$'s, as suggested by \cite{1993ApJ...415..715G} and \cite{2018arXiv180406920E}. Such a behaviour is apparent for the Plummer case and less evident in the Hernquist case. However, even considering smaller softening lengths than those of the data shown in Fig.\ \ref{figlmaxself}, we do not find clear evidences of a threshold  $\epsilon_c$ between constant and decreasing $\Lambda(N)$ depending on $N$ as $\epsilon_c\sim r_c/\sqrt{N}$.\\
\indent As a general trend, in the Plummer case the $N$-dependence of $\Lambda_{\rm max}$ at large $N$ is well described by a $N^{-1/2}$ law (solid lines in Fig. \ \ref{figlmaxself}), while in the Hernquist case the $N$-dependence is weaker and, although the values are consistently decreasing up to $N \approx 10^5$, a saturation for large values of $N$ or a matching with a $N^{-1/3}$ law (dashed lines in Fig. \ \ref{figlmaxself}) can not be totally ruled out by our results. We recall that a $\Lambda_{\rm max}\propto N^{-1/3}$ behaviour for softening lengths larger than the threshold $\epsilon_c$ was estimated by \cite{1993ApJ...415..715G} and also by \cite{1986A&A...160..203G}, in the latter case using a differential geometry approach to the gravitational $N-$body problem.\\
\indent We note that our results for $\Lambda_{\rm max}$ are given in units of $t^{-1}_{\rm dyn}$, that, as mentioned above, is always equal to unity.
\cite{2003Ap&SS.283..347C} introduced a dimensionless maximal Lyapunov exponent or ``dimensionless chaoticity indicator'' as $\gamma_1\equiv\Lambda_{\rm max}t_{\rm dyn}$. Such quantity, at variance with $\Lambda_{\rm max}$ computed here, was found to be independent of $N$ and on the average particle energy. Note that, however, the value of $t_{\rm dyn} = \sqrt{G\rho}$ used by \cite{2003Ap&SS.283..347C} depends on the mass density $\rho$ that we keep fixed, while in their case varies as a function of $N$, thus restoring a dependence on the number of particles of the dynamical time. 
\section{Summary and conclusions}
\label{summary}
We have investigated the chaotic properties of the dynamics of $N$-body self-gravitating systems, studying single-particle orbits in frozen and active potentials as well as the dynamics in the full $6N$-dimensional phase space. Initial conditions were always drawn from spherically symmetric stationary models, namely, the flat-cored Plummer model and the centrally cuspy Hernquist model. The main results of this study can be summarized as follows.\\
\indent  
As to the properties of single-particle orbits, the orbital structure of particles in frozen $N$-body models approaches that of continuum potentials, however the dependence on $N$ of this trend is not trivial. Moreover, orbits in frozen models and active self-consistent models have (obviously) different mixing properties and have, typically, different largest Lyapunov exponents; the dependence of the largest Lyapunov exponent on $N$ is more pronounced in frozen systems than in active ones.\\
\indent In general, the differences between an orbit propagated in a frozen model or in a live $N$-body model and their smooth potential counterpart become evident already after a few dynamical times, therefore on a scale much smaller than the collisional relaxation time $t_{2b}$. For example, for the model with $N=40000$, the integration time of 200$t_{\rm dyn}$ is roughly equal to $0.1t_{2b}$, ruling out the dynamical collisions  (in the sense of Chandrasekhar theory) as the origin of the differences with the analytical orbits in the smooth potential. We therefore conjecture that the nature of the fluctuations in the full N-body potential, and possibly their scales and space time correlations, induces an effect on the orbital evolution of particles that challenges the picture of independent, additive, local impulse perturbations assumed in the diffusion process underlying the Chandrasekhar description. Remarkably, at fixed $N$ the differences between orbits in frozen and live $N$-body potentials depend on the specific orbital parameters at hand. In general, orbits with lower initial angular momentum evolve more in the live case, where an excursion in both $J$ and $\mathcal{E}$ is possible.\\
\indent Given that orbits in potentials that are integrable in the continuum limit behave differently in frozen and self-consistent simulations, we argue that nothing can be safely deduced from studies with tracer particles in frozen $N$-body models reproducing non-integrable models, such as, for example, triaxial systems.\\  
\indent As far as the chaotic properties of the dynamics in the full $6N$-dimensional phase space is concerned, our results confirm the expectation (and previous results, at least qualitatively; see e.g. \citealt{2002MNRAS.331...23E}) that systems with larger $N$ are less chaotic. However, the actual value of the largest Lyapunov exponent as well as its dependence on $N$ depend on the chosen equilibrium model: the flat-cored Plummer model has smaller Lyapunov exponents that are proportional to $N^{-1/2}$, while the Lyapunov exponents of the cuspy Hernquist model are systematically larger than the Plummer ones and their dependence on $N$ is considerably weaker. The values of the largest Lyapunov exponent obviously depend on the softening length, given that systems with different softenings are different dynamical systems; as expected, smaller softening lengths yield larger Lyapunov exponents, i.e., more chaotic systems. This notwithstanding, the Lyapunov exponent does decrease with increasing $N$, at least for sufficiently large $N$'s. The dynamical behaviour is thus consistent with a regular (non-chaotic) behaviour in the continuum ($N\to\infty$) limit, although no numerical experiment can obviously prove that unambiguously.\\ 
\indent It is important to stress that we considered only systems with a finite softening length, being interested in investigating the relation with the collisionless limit, that requires a finite softening length. Our results cannot be extrapolated to the unsoftened case, hence are not in contradiction with previous results on the degree of chaoticity by \cite{1993ApJ...415..715G}, \cite{2002ApJ...580..606H} and \cite{2018arXiv180406920E}, that seem to suggest a $\Lambda_{\rm max}$ independent of $N$ for the unsoftened gravitational $N-$body problem.  We note that the r\^{o}le played by softening in affecting chaos in self-gravitating systems had been also studied by \cite{2009PhyA..388..639K}, that found a linear decrease of $\Lambda_{\rm max}$ with increasing $\epsilon_{\rm soft}$ at fixed $N$.\\
\indent This established, it remains to clarify whether the present results could be extended to axisymmetric and triaxial models (admitting in the continuum limit the coexistence of regions of regular and chaotic dynamics), or spherical models with velocity anisotropy. The interplay between $N$-body chaos and external noise in models characterized by Osipkov-Merritt radial anisotropy  will be explored in a forthcoming publication.  
\section*{Acknowledgments}
We thank Matteo Sala, Dominique Escande and  Francesco Ginelli for insightful discussions and relevant comments at an early stage of this work. The anonymous Referee is also warmly acknowledged for her/his important suggestions that have helped improving the presentation of our results. One of us (PFDC) wishes to thank Christos Efthymiopoulos and the hospitality of the Center for Astronomy and Applied Mathematics of the Academy of Athens where part of this work was done.
\bibliographystyle{mnras}
\bibliography{biblio.bib}
\end{document}